# Modelling the mechanism of cell inactivation by light ions at different energy values

Pavel Kundrát[*], Hana Hromčíková, Miloš Lokajíček
Institute of Physics, Academy of Sciences of the Czech Republic,
Na Slovance 2, CZ-182 21 Prague 8

**Abstract**

For efficient application of protons and light ions in radiotherapy, detailed knowledge and realistic models of the corresponding radiobiological mechanism are necessary. Basic characteristics of this mechanism have been represented within a probabilistic two-stage model. The processes that occur immediately after the traversals of individual particles and the response of cell to the total damage formed by all the particles have been distinguished. The model involves a probabilistic description of DNA damage formation and repair processes, too.

**Keywords:** survival curves, cell inactivation, radiobiological mechanism, mathematical models
*Short title: Model of cell inactivation by light ions*

## 1. Introduction

Accelerated protons and light ions have already proved their usefulness in clinical radiotherapy. To take the advantage of their possible benefits and to optimize treatment procedures in individual cases, the detailed understanding of underlying radiobiological mechanism is necessary. An important role is played by mathematical modelling that allows using optimization procedures.

Until now the models proposed earlier for conventional radiation, e.g. the linear-quadratic model (LQ model), have been used in describing survival curves for proton and ion irradiation, too. The corresponding model parameters, e.g. $\alpha$ and $\beta$ in the LQ model, have been derived on a phenomenological basis.

The first systematic model approach that has aimed at evaluating the biological effectiveness of light ion beams has been proposed by Kraft and Scholz [9, 3]. Their local effect model (LEM) has enabled to represent the differences in biological efficiency of diverse particles, which have been related to differences in their track structures. However, the actual underlying biological processes have not been addressed by the LEM scheme.

In the contradistinction, the probabilistic two-stage model, used in the present paper, aims at describing cell inactivation in a more realistic way by representing basic characteristics of the main processes involved in the given mechanism. In comparison to phenomenological models, the two-stage model enables to represent systematically not only the global shape of cell survival curves but also their detailed structure. Deviations from their parabolic shape in the clinically relevant region, found experimentally (compare e.g. [1, 8]), have been related to detailed characteristics of the given mechanism.

## 2. Probabilistic two-stage model

---

[*] Corresponding author. Institute of Physics, Academy of Sciences of the Czech Republic, Na Slovance 2, CZ-182 21 Prague 8. E-mail: Pavel.Kundrat@fzu.cz, Phone +420 266 052 665, Fax +420 286 585 443.

*Basic scheme*

The radiobiological mechanism may be divided into two different stages: (i) energy transfer from individual ionizing particles to a cell and formation of chemical damage of important biomolecules (physical and chemical phases), and (ii) the reaction of individual cells to the damages caused by different numbers of energy transfers, their frequency distribution corresponding to applied dose. These two steps have been represented by the probabilistic two-stage model [2, 4-6]. The model scheme applicable to the case of irradiating by monoenergetic ions will be described in the following.

The average number $k_{av}$ of primary particles traversing chromosomal system of the cell (energy transfer events) is given by applied dose, $D$, and number of particles per unit dose, $h$:

$$k_{av} = hD = C \sigma D / \lambda, \tag{1}$$

where $\sigma$ stands for effective geometrical cross-section of the chromosomal system[†] and $\lambda$ represents linear energy transfer (LET); conversion constant $C$ = 6.24 keV/Gy/µm$^3$.

The actual numbers $k$ of primary particles traversing chromosomal systems in individual cells are of stochastic nature. Supposing that impacts of individual particles are fully random, the distribution of traversing particle numbers may be described by Poisson statistics,

$$P_k = \exp(-hD) \cdot (hD)^k / k! \; . \tag{2}$$

Distribution of individual energy transfer events and the amount of transferred energy (for monoenergetic particles given by LET value $\lambda$) stand for the primary characteristics of the physical phase. As to the subsequent processes, let us denote by $p_k(\lambda)$ the average probability of cell inactivation after the impact of $k$ primary particles of LET value $\lambda$; detailed model of these parameters will be discussed later on. The probability of a cell to survive after dose $D$ can be written, then, as

$$s(D) = 1 - \Sigma_k P_k(D) p_k(\lambda) \; . \tag{3}$$

Inactivation probabilities $p_k$ after different numbers of hits represent basic parameters involved in the two-stage radiobiological model. They possess direct physical and biological interpretation, and represent a basis for detailed modelling of the radiobiological mechanism. Their values can be derived by analyzing experimental cell survival data, either using an auxiliary polynomial fitting procedure [2, 4] or directly using Eqs. (1-3), which is free of numerical difficulties met by the former method (compare [4, 7]). More detailed mechanistic model, relating inactivation effect to the processes of DNA damage induction and repair, will be described in the following paragraph.

*Inactivation probabilities $p_k$*

In the case of protons and light ions, even a single particle may form a damage that results finally in inactivating the cell. Let us denote by $a(\lambda)$ the probability that such a damage has been induced, per 1 traversing particle. The

---

[†] Note that $\sigma$ [µm$^2$] denotes the geometrical effective cross-section of the chromosomal system (or of sensitive region within cell nucleus) to be traversed by the given particles. The consequent effects of DNA damage and cellular response are treated separately. Note especially that $\sigma$ is not identical to cell inactivation cross-section, which is used by some authors to describe (the linear component of) the probability, per 1 particle, of the cell to be inactivated; compare e.g. [9].

probability that no such damage has been formed after the impact of $k$ particles is given by

$$q_k^A = (1 - a(\lambda))^k. \tag{4}$$

Traversing particles might induce sublethal damages, too, that may combine and form lethal ones. Let us denote by $b(\lambda)$ the probability that one particle forms a corresponding sublethal damage. The probability that no combination of sublethal damages occurred after the traversal of $k$ particles is then given by

$$q_k' = (1 - b^2(\lambda))^{k(k-1)/2}, \tag{5}$$

where we have assumed, for the sake of simplicity, that the synergetic effect arises mainly from the combination of two such damages. A part of these combined damages is assumed to be repaired by the cell. Let us denote the probability of their successful repair by $r(\lambda,k)$. The probability that the cell is not inactivated by the considered combination of individual sublethal damages equals then

$$q_k^B = 1 - (1 - q_k').(1 - r(\lambda,k)). \tag{6}$$

When both damage types are taken into account, the formula for cell survival probability after the impact of $k$ particles reads

$$p_k = 1 - q_k^A \cdot q_k^B. \tag{7}$$

The cell survival probability after applied dose $D$ is obtained, then, by convolving $p_k$ with Poisson distribution of traversing particles, $P_k$, according to Eqs. (1-3).

The functions $a(\lambda)$, $b(\lambda)$ and $r(\lambda,k)$ have to be established in agreement with corresponding experimental data (compare Sec. 3). More detailed microscopic models might help in their evaluation, too.

## 3. Analysis of experimental data

The probabilistic two-stage model described in the preceding has been applied to experimental data obtained by Belli et al. [1] in irradiating Chinese hamster V79 cells by monoenergetic protons. We have analysed six survival curves with LET values ranging from 7.7 to 37.8 keV/μm.

To obtain precise fits of the data, the following flexible smooth test functions with low number of auxiliary free parameters have been used for $a(\lambda)$, $b(\lambda)$, $r(\lambda,k)$ in the given analysis:

$$a(\lambda) = (a_1\lambda + a_2\lambda^2)\,[1-\exp(-(a_3\lambda)^{a_4})],$$

$$b(\lambda) = [1-\exp(-(b_1\lambda)^{b_2})] / [1+b_3\exp(-(b_4\lambda)^{b_5})], \tag{8}$$

$$r(\lambda,k) = 1-[1-\exp(-(r_1\lambda k)^{r_2})] / [1+r_3\exp(-(r_4\lambda k)^{r_5})].$$

The values of the auxiliary parameters were found by numerical optimization procedure:
  $a_{1-4}$: 0.002, 0.013, 0.026, 5.0;
  $b_{1-5}$: 0.12, 5.0, 24.1, 0.061, 1.176;
  $r_{1-5}$: 0.024, 5.0, 0.56, 0.002, 5.0.
These values have corresponded to the effective geometrical cross-section of chromosomal system $\sigma = 12.8$ μm$^2$, also found by numerical optimization. This value is in agreement with the values of geometrical cross sections of V79 cell nuclei, $\sigma_{nucl} = 134$ μm$^2$, found in the given experiment [1].

Calculated cell survival curves together with experimental data are shown in Figure 1. The increase of the inactivation probabilities with the number of particles traversing cell chromosomal system is displayed in Figure 2.

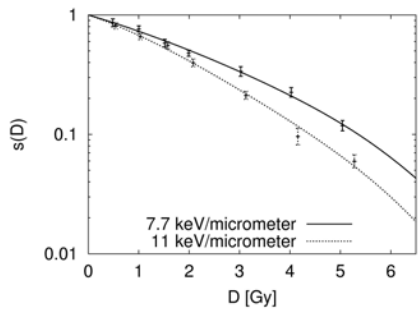
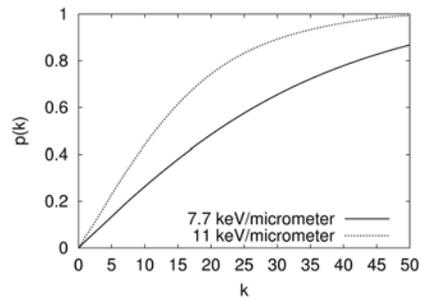
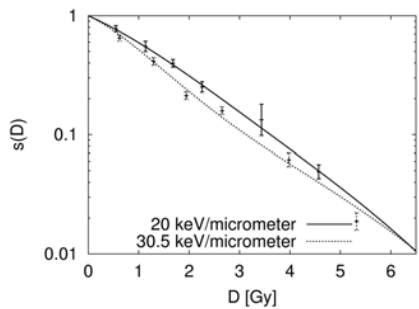
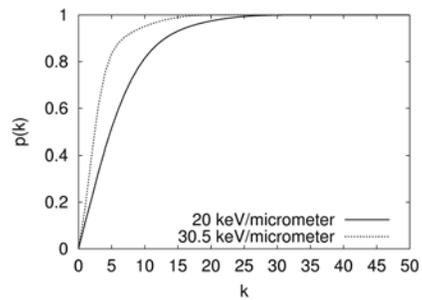
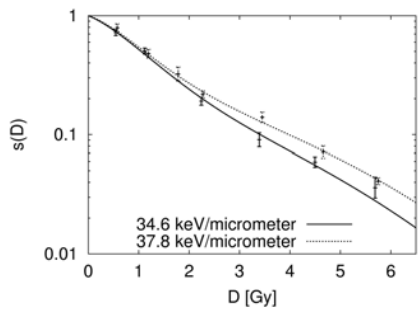
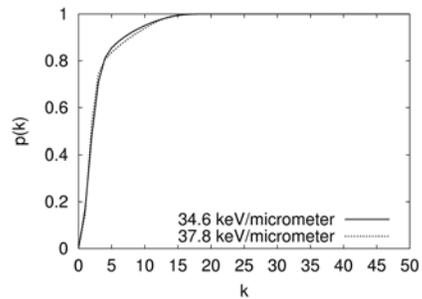

Figure 1: Survival of Chinese hamster V79 cells in proton irradiation: experimental data (taken from [1]) and their model representation

Figure 2: Inactivation probabilities $p_k$ increasing with the number of protons traversing the cell nucleus

The probability of irreparable (lethal) damage formation, $a(\lambda)$, and the probability of reparable (sublethal) damage, $b(\lambda)$, are shown in Figure 3. The probability of successful repair, $r(\lambda,k)$, is pictured in Figure 4.

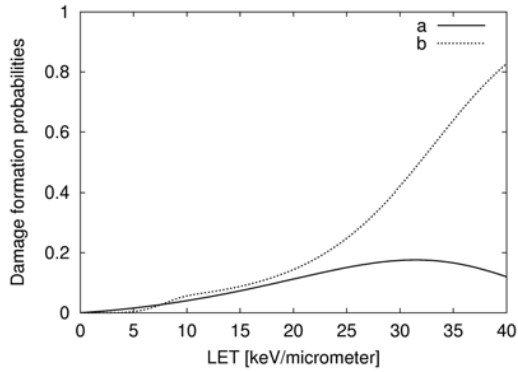 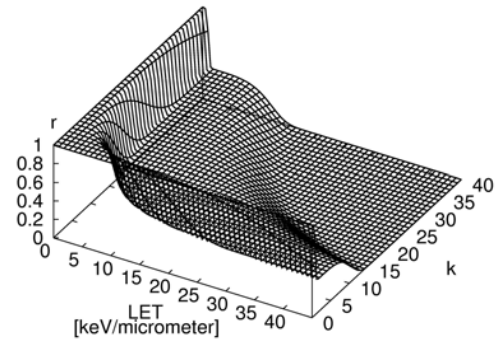

Figure 3: Probability of lethal, a(λ), and sublethal, b(λ), damage formation by individual particles

Figure 4: Probability of successful repair, r(λ,k)

## 4. Discussion

The model has enabled to represent the given set of experimental data in a consistent and precise way. Basic characteristics of DNA damage formation and of cellular repair processes have been derived for the studied case. The results indicate that the formation of lethal damages by individual protons saturates at LET values around 30 keV/μm, while the formation of sublethal damages shows a steady increase over the whole studied range of proton LET values. The probability of successful repair shows reasonable behaviour, too. The decreasing repair probability with increasing number of particles and/or their LET values corresponds to the increase in complexity of the DNA damage formed under such conditions.

The precise parameterization of damage induction and repair, Eq. (8), as well as the parameters involved, play merely an auxiliary role. Realistic interpretation can be attributed to damage induction and repair probabilities only. If a less precise description of the data is sufficient, test functions involving a lower number of free parameters might be used instead of those of Eq. (8); compare [5-7].

Similar analyses have been performed for cell inactivation by other ions, too, showing the feasibility of applying the given model and the possibility of quantifying the roles of DNA damage formation and repair by this approach; see [5-7]. Saturation effects in damage induction, as discussed above for protons, have not been observed for other ions, indicating a certain difference in the detailed mechanism of effects between protons and heavier ions.

## 5. Conclusion

The probabilistic two-stage model provides a realistic description of the radiobiological mechanism. It includes description of irreparable and reparable damages formed by individual particles. Basic characteristics of repair processes are incorporated, too. The given model might serve as a basis for more detailed microscopic modelling of radiobiological effects.

The model enables to represent not only the global shape of experimental cell survival curves but also their detailed behaviour. This is necessary if different fractionation schemes in clinical hadron radiotherapy are to be evaluated, as even small local deviations are largely amplified in such a case.